\documentclass[preprints,article,accept,moreauthors,pdftex]{arxivdoc}

\firstpage{1} 
\pubvolume{1}
\issuenum{1}
\articlenumber{0}
\pubyear{2021}
\copyrightyear{2020}
\datereceived{} 
\dateaccepted{} 
\datepublished{} 
\hreflink{https://doi.org/} 

\usepackage{hyperref}
\usepackage{upgreek}

\Title{Long-range atom-ion Rydberg molecule: A novel molecular binding mechanism}

\TitleCitation{Long-range atom-ion Rydberg molecule: A novel molecular binding mechanism}

\Author{Markus Dei{\ss} $^{1}$, Shinsuke Haze $^{1}$ and Johannes Hecker Denschlag $^{1,}$*{}}

\AuthorNames{Markus Dei{\ss}, Shinsuke Haze and Johannes Hecker Denschlag}

\AuthorCitation{Dei{\ss}, M.; Haze, S.; Hecker Denschlag, J.}

\address{%
$^{1}$ \quad Institut f\"{u}r Quantenmaterie and Center for Integrated Quantum Science and Technology IQ$^{ST}$, Universit\"{a}t Ulm, 89069 Ulm, Germany; johannes.denschlag@uni-ulm.de}

\corres{Correspondence: johannes.denschlag@uni-ulm.de}

\abstract{We present a novel binding mechanism where a neutral Rydberg atom and an atomic ion form a molecular bound state at large internuclear distance. The binding mechanism is based on Stark shifts and level crossings which are induced in the Rydberg atom due to the electric field of the ion. At particular internuclear distances between Rydberg atom and ion, potential wells occur which can hold atom-ion molecular bound states. Apart from the binding mechanism we describe important properties of the long-range atom-ion Rydberg molecule, such as its lifetime and decay paths, its vibrational and rotational structure, and its large dipole moment. Furthermore, we discuss methods how to produce and detect it. The unusual properties of the long-range atom-ion Rydberg molecule give rise to interesting prospects for studies of wave packet dynamics in engineered potential energy landscapes. 
}


\begin{document}
	
\section{Introduction}
	
Molecules or bound complexes are often classified according to their binding mechanisms. The covalent bond, the ionic bond, the van der Waals bond, and the hydrogen bond are perhaps the ones that are most widely known, but they represent only a selection of all possible kinds of bonds. A recently discovered class of molecular bound states are long-range Rydberg molecules between two or more neutral atoms, where binding lengths can be in the micrometer range. After their prediction about two decades ago \cite{Greene2000}, several types of long-range Rydberg molecules have been experimentally observed in recent years \cite{Bendkowsky2009, Booth2015, Niederpruem2016, Sassmannshausen2016, Hollerith2019} (for reviews, see, e.g., \cite{Shaffer2018, Fey2019, Eiles2019}). 
	
In this work, we predict another species of the long-range Rydberg molecules. Here, a Rydberg atom is bound at a large given distance to an ion. The electric field of the ion leads to Stark shifts of the  energy levels of the Rydberg atom, which strongly depend on the relative distance to the ion. At positions, where avoided crossings between low-field-seeking and high-field-seeking energy levels occur, potential wells exist which exhibit molecular bound states.
	
In a simple classical picture the bound state is based on the electrostatic interaction between an electric dipole and a charge. Depending on the orientation of the electric dipole the interaction can be attractive or repulsive. The electric field of the ion generally polarizes the neutral Rydberg atom according to its polarizability. This polarizability is a function of the electric field that the atom is exposed to and thus changes with the distance between atom and ion. The idea is now that a  long-range Rydberg molecule can form at a distance where the polarizability flips sign, such that at shorter distances there is repulsion between atom and ion and at larger distances there is attraction. 
	
As we show, a Rydberg series of these atom-ion long-range bound states exists, within which the binding energies and bond lengths strongly vary. We discuss properties of the novel molecules such as their stability and lifetime as well as their rotational and vibrational characteristics. Due to their large electric dipole moment they can easily be aligned in a weak electric field. We propose controlled coupling of various quantum states via radio-frequency or microwave radiation. Finally, we describe how the predicted molecules can be created and detected in a cold atom experimental setup.

Before we start our investigation of the long-range atom-ion Rydberg bound states we would like to mention, that in recent years there has been a growing interest in the collisions and interactions of cold atoms and ions (for reviews, see, e.g., \cite{Haerter2014,Tomza2017}). As of late, also the interactions between ions and ultracold Rydberg atoms have been studied theoretically \cite{Secker2016, Hirzler2021, Secker2017, Wang2020} as well as experimentally \cite{Ewald2019, Haze2019, Engel2018, Gross2020}. Building up on this, an observation of the proposed long-range atom-ion Rydberg molecule may be possible in the near future.


\section{Binding mechanism and properties of the long-range atom-ion Rydberg molecules}
	
In the following we explain in detail the binding mechanism of the long-range atom-ion molecule. It consists of a neutral Rydberg atom and an ion which are at a large enough internuclear distance $r$ so that there is negligible overlap of the Rydberg electron orbital with the ion. We assume, for now, that the ionic core of the neutral atom is located at the origin. The ion is located on the $z$-axis at $\vec{r} = (x = 0,\,y = 0,\,z = r) $. For simplicity, we consider the ion to be a point charge, with a single, positive elementary charge $e$. The ion generates a spherically symmetric electric field of strength $\mathcal{E} =e/(4\pi\varepsilon_0|\vec{r} - \vec{x} |^2)$ at  location $\vec{x} $. Here, $\varepsilon_0$ is the vacuum permittivity. The electric field of the ion leads to level shifts and crossings in the atomic Rydberg atom, based on the Stark effect. The resulting level structures are quite similar to the well-known Stark maps of a Rydberg atom in a homogeneous electric field.

\begin{figure}[t]
\includegraphics[width=12.6cm]{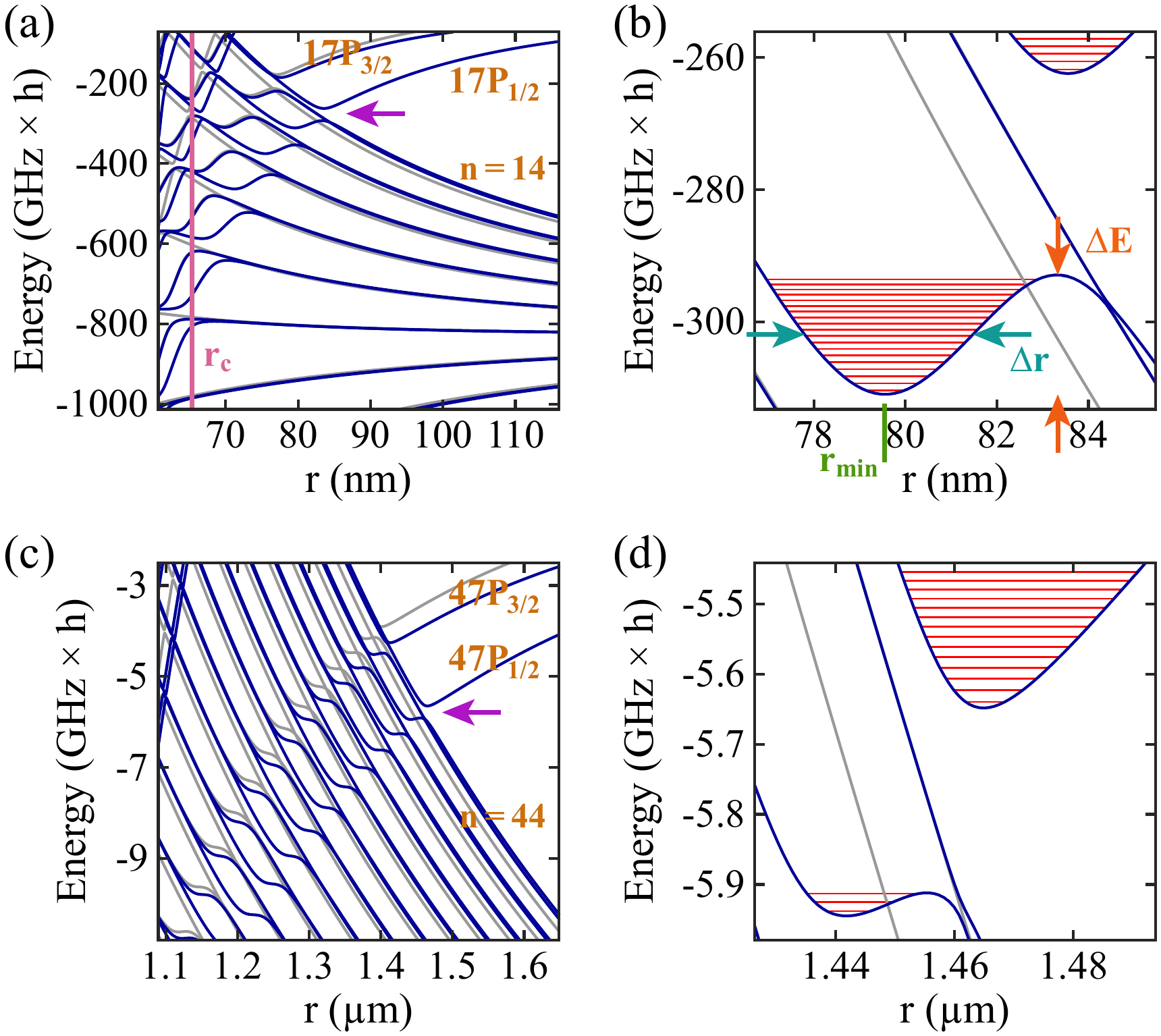}
\caption{(\textbf{a}) Stark map level structure of a Rb atom in the vicinity of the $17P$ state as a function of the internuclear distance $r$ between the atom and the ion. The energy reference is given by the term energy of the $17P_{3/2}$ state at zero electric field corresponding to the value for $r\rightarrow \infty$. Blue (gray) solid lines represent levels with $|m_J|=1/2$ ($|m_J|=3/2$), where $m_J$ is the magnetic quantum number. The purple vertical line marks a critical internuclear distance $r_\mathrm{c}$. For $r \lesssim r_\mathrm{c}$ our model breaks down (see text). (\textbf{b}) Zoom into the region indicated by the magenta arrow in Figure~\ref{Fig1}a, showing the two outermost potential wells. The red solid horizontal lines correspond to vibrational level energies. The parameters $\Delta E$, $\Delta r$, and $r_\mathrm{min}$ are used to characterize a potential well (see text). (\textbf{c} and \textbf{d}) Stark map and zoom in the vicinity of the $47P$ state. Here, the energy reference is the term energy of the $47P_{3/2}$ state at zero electric field. \label{Fig1}
}
\end{figure}

For the sake of a concrete example, we choose Rb as the neutral atom species. Figure~\ref{Fig1} shows Rydberg levels of a Rb atom as a function of the internuclear distance $r$ between atom and ion. Specifically, the level structures in the vicinity of $nP$ states are considered. The principal quantum number is $n=17$ for Figures~\ref{Fig1}a and \ref{Fig1}b, while it is $n=47$ for Figures~\ref{Fig1}c and \ref{Fig1}d. In the given range of energies and internuclear distances avoided level crossings between the high-field-seeking $17P_J$ ($47P_J$) states and the low-field-seeking states of the hydrogenic manifold belonging to $n=14$ ($n=44$) start to occur. Here, $J$ denotes the total electronic angular momentum quantum number, which can be $1/2$ or $3/2$ for the $P$ states. The avoided crossings give rise to potential wells within which molecular bound states can exist. Figures~\ref{Fig1}b and \ref{Fig1}d are magnifications of the regions marked with magenta arrows in plots \ref{Fig1}a and \ref{Fig1}c. The horizontal red lines in Figure~\ref{Fig1}b (Figure~\ref{Fig1}d) are the quantum-mechanical vibrational bound state levels in the wells (for rotational angular momentum $l' = 0$). 

The physics behind the bound states is as follows. The interaction is based on the interaction between an induced dipole moment and a charge. At the location of the bottom of a potential well the dipole moment of the Rydberg atom flips its sign. At larger distances it is a high-field seeker and thus is attracted by the ion. At shorter distances it becomes a low-field seeker and is repelled by the ion. Hence, it oscillates about the bottom of the potential, the location of which corresponds to an approximate bond length.

To calculate the atomic Rydberg level energies in Figure~\ref{Fig1} we determine in a first step the unperturbed atomic Rydberg states $ | n, S, L, J, m_J \rangle $ and their energies  $E_{(n, S, L, J, m_J)}$ for Rb using the ARC (Alkali.ne Rydberg Calculator) package \cite{Robertson2021}. Here, $L$ and $S$ are the quantum numbers of the electronic orbital angular momentum and the electronic spin. These calculations include the fine structure but ignore the hyperfine structure. In a second step we take into account the electrostatic interaction between the Rydberg atom (treated as a Rydberg electron and a point-like Rydberg ionic core) and the ion,

\begin{equation}
V_I = \frac{e^2}{4\pi\varepsilon_0 \, r} - \frac{e^2}{4\pi\varepsilon_0  |\vec{r} - \vec{r}_e| }\,,
\label{eq:VI_1}
\end{equation}
where $ \vec{r}_e  $ is the location of the Rydberg electron. The potential $V_I$ can be  expressed in a multipole expansion (see, e.g., \cite{Jackson1998}) by

\begin{equation}
V_I=-\frac{e^2}{4\pi\varepsilon_0}\sum_{l=1}^{\infty}\sqrt{\frac{4\pi}{2l+1}}\frac{r_e^l}{r^{l+1}}Y_{l0}(\theta_e,\phi_e)\,,
\label{EqA1}
\end{equation}
where $l$ is the order of the multipole term and $Y_{lm}$ represent the spherical harmonics. The quantum number $m$ is  $m = 0$, due to rotational symmetry. Furthermore, the variables $(r_e, \theta_e, \phi_e)$ are the spherical coordinates for the Rydberg electron location

\begin{equation}
\vec{r}_e = r_e \, \left( \begin{array}{ccc}\sin(\theta_e)\cos(\phi_e)\\   \sin(\theta_e) \sin(\phi_e)\\ \cos(\theta_e) \\ \end{array} \right)\,.
\label{Eqcoord}
\end{equation}
We note that in Equation~\ref{EqA1} the zero order ($l = 0$) multipole term drops out, as the corresponding contributions of Rydberg electron and Rydberg ionic core cancel each other. Thus, when the electron is located at the origin, i.e. $r_e = 0$, the interaction $V_I$ vanishes. Therefore, the atomic ground state will only be comparatively weakly affected by $V_I$. 
	
The matrix $\langle \widetilde{n}, \widetilde{S}, \widetilde{L}, \widetilde{J}, \widetilde{m}_J  | V_I +  E_{(n, S, L, J, m_J)} | n, S, L, J, m_J \rangle $ represents the Hamiltonian of the static (motionless) atom-ion system for a given, fixed $r$. By diagonalizing the matrix we obtain the energy levels shown in  Figure~\ref{Fig1}. In our calculations we only take into account multipole terms up to the order of $l=6$ in Equation~\ref{EqA1}, since higher orders have negligible contributions. We note that truncating the multipole expansion of Equation~\ref{EqA1} to the lowest order $l=1$ would correspond to approximating the electric field to be homogeneous. Such a homogeneous field gives rise to the standard Stark maps \cite{Zimmerman1979}.
	
In the spirit of the Born-Oppenheimer approximation the obtained energy levels represent molecular potential energy curves of the atom-ion system. Since the atom-ion interaction potential is spherically symmetric, the angular momentum $l'$ for the molecular rotation is a good quantum number. The kinetic energy term of the rotational angular momentum gives rise to the centrifugal potential

\begin{equation}
V_\mathrm{cp}=\frac{\hbar^2 l'(l'+1)}{2\mu r^2}
\label{Eq:centrifugal}
\end{equation}
in the radial Schr\"{o}dinger equation. Here, $\mu$ denotes the reduced mass of the diatomic molecule and $\hbar = h / (2\pi)$, where $h$ is Planck's constant. Radial bound states of the resultant potential wells correspond to the vibrational eigenstates of the atom-ion molecule with angular momentum $l'$.
		
\begin{figure}[]
\includegraphics[width=13cm]{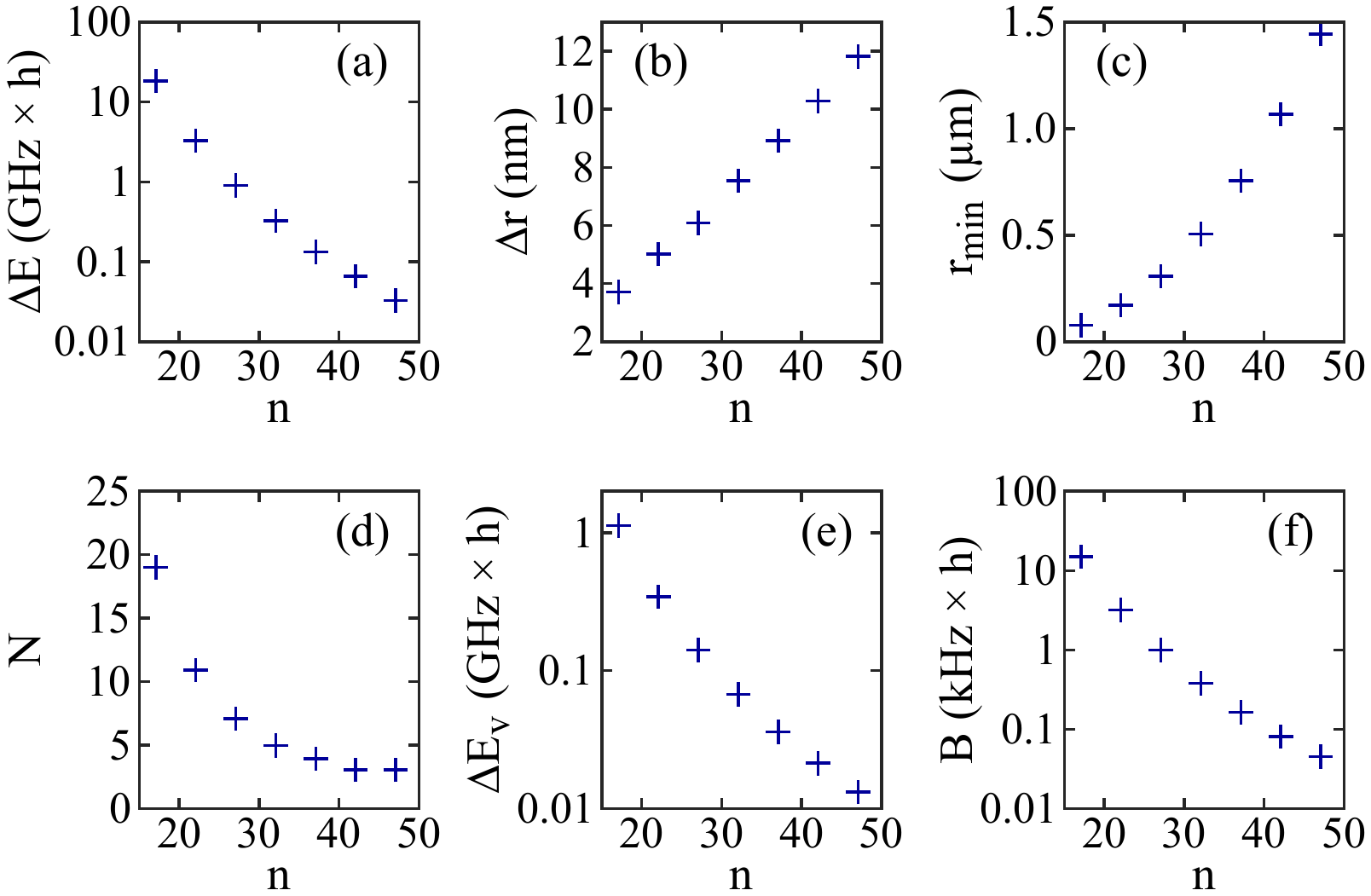}
\caption{Properties of the second outermost potential wells associated with the $nP_{1/2}$ states of Rb and of corresponding molecular bound states as functions of the principal quantum number $n$. (\textbf{a}) Depth $\Delta E$ (on a logarithmic scale). (\textbf{b}) Width $\Delta r$. (\textbf{c}) Position $r_\mathrm{min}$. (\textbf{d}) Number $N$ of vibrational bound states. (\textbf{e}) Energy splitting $\Delta E_\mathrm{v}$ between the vibrational ground state and the first excited vibrational state (on a logarithmic scale). (\textbf{f}) Rotational constant $B$ for a $^{87}$Rb$^{138}$Ba$^+$ molecule (on a logarithmic scale). \label{Fig2}
}
\end{figure}

Bound state wells, similar as the ones discussed in Figure~\ref{Fig1} also occur for different principal quantum numbers $n$. In fact, there exists a Rydberg series of potential wells for the long-range atom-ion Rydberg molecules. The well depth $\Delta E$, the width $\Delta r$ at half the depth, and the position $r_\mathrm{min}$ of the bottom of a well (see Figure~\ref{Fig1}b) change with $n$. Figure~\ref{Fig2} shows these dependencies for the second outermost potential well associated with the $nP_{1/2}$ state, as $n$ increases in steps of five units from $n=17$ to $n=47$. As can be seen in Figure~\ref{Fig2}a, $\Delta E$ decreases by about three orders of magnitude from $18\:\textrm{GHz} \times h$ for $n=17$ to about $30\:\textrm{MHz}\times h$ for $n=47$. At the same time $\Delta r$ increases from about $4\:\textrm{nm}$ to almost $12\:\textrm{nm}$ (Figure~\ref{Fig2}b), and the binding length $r_\mathrm{min}$ rises from $80\:\textrm{nm}$ to a remarkable value of $1440\:\textrm{nm}$ (Figure~\ref{Fig2}c).

As the potential wells change with $n$, so do the numbers of their vibrational bound states $N$ and the vibrational splittings of the levels. This is shown in Figures~\ref{Fig2}d to \ref{Fig2}f for the second outermost potential wells associated with the $nP_{1/2}$ states. The number of vibrational bound states $N$ decreases from $19$ to $3$ as $n$ increases from $17$ to $47$ (see Figure~\ref{Fig2}d). At the same time, the energy splitting for deeply bound vibrational levels drops from $~1\:\textrm{GHz}\times h$ to $~10\:\textrm{MHz}\times h$ (see Figure~\ref{Fig2}e). As the atom-ion interaction potential is spherically symmetric it has rotational eigenstates. The rotational constant $B$ can be estimated in the approximation of a rigid rotor using $B=\hbar^2/(2 \mu r_\mathrm{min}^2)$, where $r_\mathrm{min}$ represents the binding length. Generally, $B$ is quite small due to the long-range character of the dimer. For example, in Figure~\ref{Fig2}f we show the rotational constant for a $^{87}$Rb atom bound to a $^{138}$Ba$^+$ ion as a function of $n$ for the given second outermost potential well. $B$ decreases from $15\:\textrm{kHz} \times h$ for $n=17$ to $50\:\textrm{Hz} \times h$ for $n=47$.
	
\section{Stability and lifetime of the long-range atom-ion Rydberg molecules}
	
One possible decay channel for a long-range atom-ion Rydberg molecule is radiative decay. The radiative lifetime of a molecular state will be generally similar as the one for the corresponding atomic Rydberg state. Considering decay due to spontaneous photon emission and assuming zero temperature, the lifetime $\tau$ of Rydberg atoms increases with $n_\mathrm{eff}$ as $\propto n_\mathrm{eff}^\alpha$ \cite{Gallagher1994, Saffman2010, Loew2012}, where $\alpha \approx 3$ for the alkali atoms. Here, $n_\mathrm{eff}=n-\delta (n)$ is the effective principal quantum number and $\delta (n)$ represents the quantum defect. At finite temperature $T$ the total lifetime $\tau_T$ can be obtained from $1/\tau_T=1/\tau+1/\tau_\mathrm{bb}$, where $1/\tau_\mathrm{bb}$ is the decay rate due to black-body radiation. We note that $\tau_\mathrm{bb}$ approximately follows the scaling law $\tau_\mathrm{bb} \propto n^2/T$ for large $n$. For example, the lifetimes of the $17P_{1/2}$ and $17P_{3/2}$ levels of $^{87}$Rb are about $5\:\upmu \textrm{s}$ ($8\:\upmu \textrm{s}$) at $T=350\:\textrm{K}$ ($T=0\:\textrm{K}$) according to the calculation in \cite{Theodosiou1984}, which includes the effects due to the core polarizability, spin-orbit interaction, and black-body radiation. A lifetime of $5\:\upmu \textrm{s}$ corresponds to a natural linewidth of $200\:\textrm{kHz}/(2\pi)$. This is, by the way, already larger than $B/h$ for the rotational constant of $15\:\textrm{kHz}\times h$ we calculated in the previous section. Therefore, low rotational levels within a vibrational state cannot be resolved.
	
Another possible decay channel is tunneling of the molecule through the outer (or inner) potential barrier of a well. Let us consider the second outermost potential well associated with the $17P_{1/2}$ state, and located at $r_\mathrm{min} \approx 80\:\textrm{nm}$ (see Figures~\ref{Fig1}a and \ref{Fig1}b). Here, the most weakly bound vibrational states may undergo tunneling towards larger internuclear distances. After passing the barrier the molecule accelerates along the repulsive potential energy curve and dissociates. The transmission probability $P_t$ of a molecule impinging on a barrier can be estimated by

\begin{equation} P_t=\mathrm{exp}\left(-2\frac{\sqrt{2\mu}}{\hbar}\int\limits_{T_{p_1}}^{T_{p_2}}\sqrt{V(r)-E_m}\,dr\right)\,,
\end{equation}
where $E_m$ is the energy of the molecule.  $T_{p_1}$ and $T_{p_2}$ are the classical turning points for the potential barrier $V(r)$. For the given well we obtain $P_t = 2.7\times 10^{-2}$, $P_t =5.9\times 10^{-4}$, $P_t=1.1\times 10^{-5}$, and $P_t=2.4\times 10^{-7}$ for the four energetically highest vibrational levels, respectively. As expected, there is a fast increase of $P_t$ with vibrational quantum number close to the threshold of the barrier. A molecular decay rate $\gamma_t$ due to tunneling can be estimated by multiplying the tunneling probability with the frequency of the oscillatory motion, which is about $1\:\textrm{GHz}$. Therefore, $\gamma_t$ is on the order of a few $10^7\:\textrm{s}^{-1}$ for the most weakly bound level. For tunneling out of the corresponding well associated with the $47P_{1/2}$ state (see Figures~\ref{Fig1}c and \ref{Fig1}d) the calculation yields $P_t=0.38$, $P_t=3.3\times 10^{-3}$, and $P_t=4.8\times 10^{-5}$ for the three available vibrational states. The frequency of the oscillatory motion, however, is only about $10\:\textrm{MHz}$ and therefore, the decay rates $\gamma_t$ are still moderate. 
	
In principle, a long-range atom-ion Rydberg molecule can also decay at the bottom of its potential well due to a nonadiabatic transition to another potential energy curve. Within the Landau-Zener theory \cite{Landau1932, Zener1932} for avoided crossings the probability for nonadiabatic transfer is given by

\begin{equation}
P_\mathrm{LZ}=\mathrm{exp}\left( -\frac{\pi \hbar \Omega^2}{2v_\mathrm{a}\,\frac{dE(r)}{dr}} \right)\,,
\end{equation}
where $\hbar \Omega$ is the energy splitting at the crossing, $v_\mathrm{a}$ is the velocity for the approach to the energy gap, and $\frac{dE(r)}{dr}$ is the differential slope of the two crossing potential energy curves. For the potential wells we consider here, we find, however, that the probabilities for nonadiabatic Landau-Zener transitions are completely negligible. 

\begin{figure}[t]
\includegraphics[width=13cm]{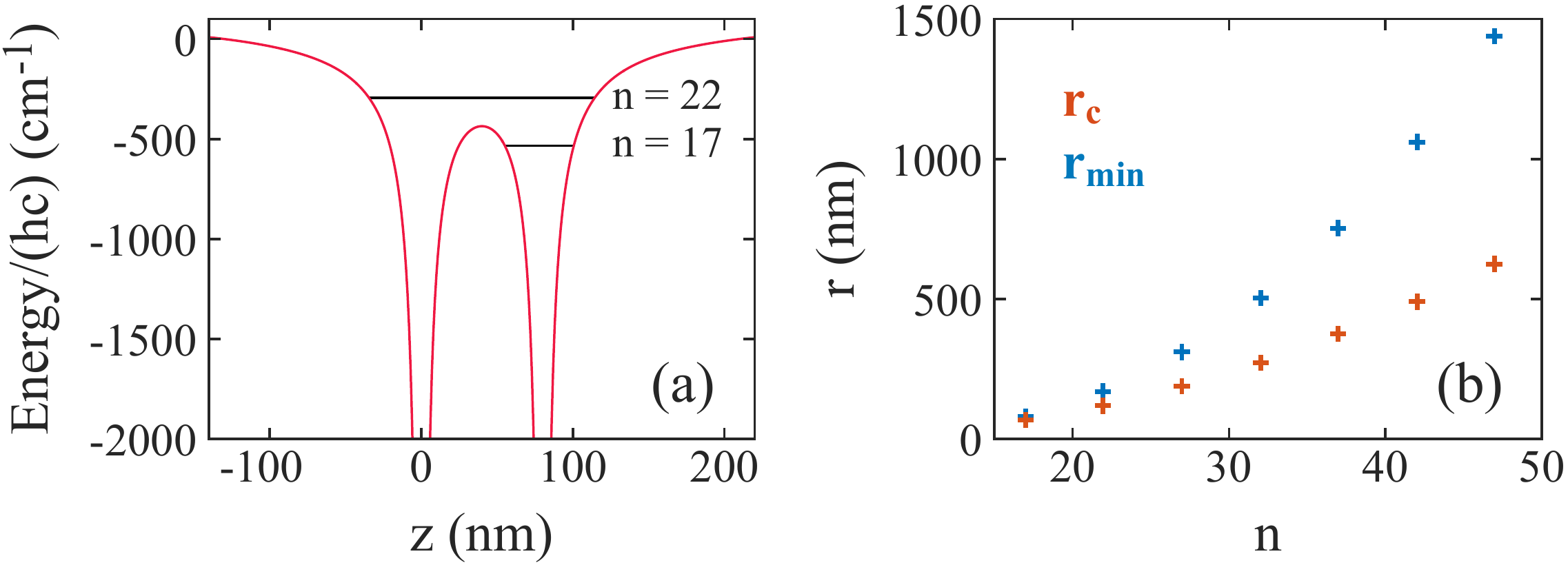}
\caption{(\textbf{a}) Over barrier motion. The red solid lines represent the Coulomb potential $V_\mathrm{Cou}$ as given in Equation~\ref{EqCoulomb} for two ionic cores located at $z=0$ and $z=80\:\textrm{nm}$ on the $z$-axis (i.e., $r=80\:\textrm{nm}$) and an electron at position $z$. The black solid horizontal lines correspond to the energy of the unperturbed Rydberg level for $n=17$ and $n=22$, respectively. (\textbf{b}) Comparison of the critical internuclear distance $r_\mathrm{c}$ to the internuclear distance $r_\mathrm{min}$ where the minimum of the second outermost potential well associated with $nP_{1/2}$ states is located, as a function of $n$. \label{Fig3}
}
\end{figure}
	
Finally, we discuss the stability of long-range atom-ion Rydberg molecules in terms of over barrier motion \cite{Ostrovsky1995}. In Figure~\ref{Fig3}a the red solid lines show the total Coulomb potential of the Rydberg electron and the two ionic cores

\begin{equation}
V_\mathrm{Cou} = \frac{e^2}{4\pi \varepsilon_0} \left(  - \frac{1} {|z|}   -  \frac{1}{|r - z|}   +   \frac{1}{r}  \right)\,,
\label{EqCoulomb}
\end{equation}
where all three particles are located on the $z$-axis. The electron is at position $z$ and the ionic cores are at positions  $z=0$ and $z=80\:\textrm{nm}$ (such that their distance is $r = 80\:\textrm{nm}$). At half the distance between the cores, i.e. at $z=40\:\textrm{nm}$, a potential barrier for the electron occurs. Our type of long-range atom-ion Rydberg molecule cannot energetically exist above this potential barrier because the electron could freely pass from one ionic core to the other one, corresponding to charge exchange between atom and ion. There still might be molecular bound states in this regime, but these are of a different type and we will not consider them any further here. We note, that even for energies slightly below the potential barrier charge exchange and a breakdown of our scheme might occur, due to tunneling of the electron through the barrier.
	
In the following we roughly estimate at what internuclear distance over barrier motion will set in if the initial atomic Rydberg state is a $nP_{1/2}$ state. For this, we simply compare the energy of an electron in an unperturbed atomic Rydberg level to the barrier height. The energy of an unperturbed Rydberg level is $-Ry\,n_\mathrm{eff}^{-2}$, where $Ry \approx 13.605\:\textrm{eV}$ is the Rydberg energy. For calculating the quantum defect $\delta (n)=\delta_0+\delta_2/(n-\delta_0)^2$ for $nP_{1/2}$ states of Rb we use the quantum defect parameters $\delta_0=2.6548849(10)$ and $\delta_2=0.2900(6)$ from \cite{Li2003}. The barrier energy at the top is $ -  3 e^2/(\pi \varepsilon_0 r )$. Therefore, the critical distance below which over the barrier motion occurs is given by

\begin{equation}
r_\mathrm{c} = \frac{3 e^2}{4 \pi \varepsilon_0  Ry} \,  (n - \delta(n))^{2}\,.
\end{equation}
This is plotted in Figure~\ref{Fig3}b along with values $r_\mathrm{min}$ for the locations of the second outermost potential wells associated with the $nP_{1/2}$ states. For large $n$, $r_\mathrm{c}$ is significantly smaller than $r_\mathrm{min}$ and therefore the potential barrier prevents charge exchange, protecting the long-range atom-ion Rydberg molecules. As $n$ decreases, $r_\mathrm{c}$ approaches  $r_\mathrm{min}$. For the lowest $n$ considered here, $n=17$, the calculation yields $r_\mathrm{c}=65\:\textrm{nm}$, which is already close to $r_\textrm{min} \approx 80\:\textrm{nm}$ (see also Figure~\ref{Fig1}a).
	
\section{Production and detection of the long-range atom-ion Rydberg molecules}

\subsection{Production by photoassociation}
	
\begin{figure}[t!]
	\includegraphics[width=12.8cm]{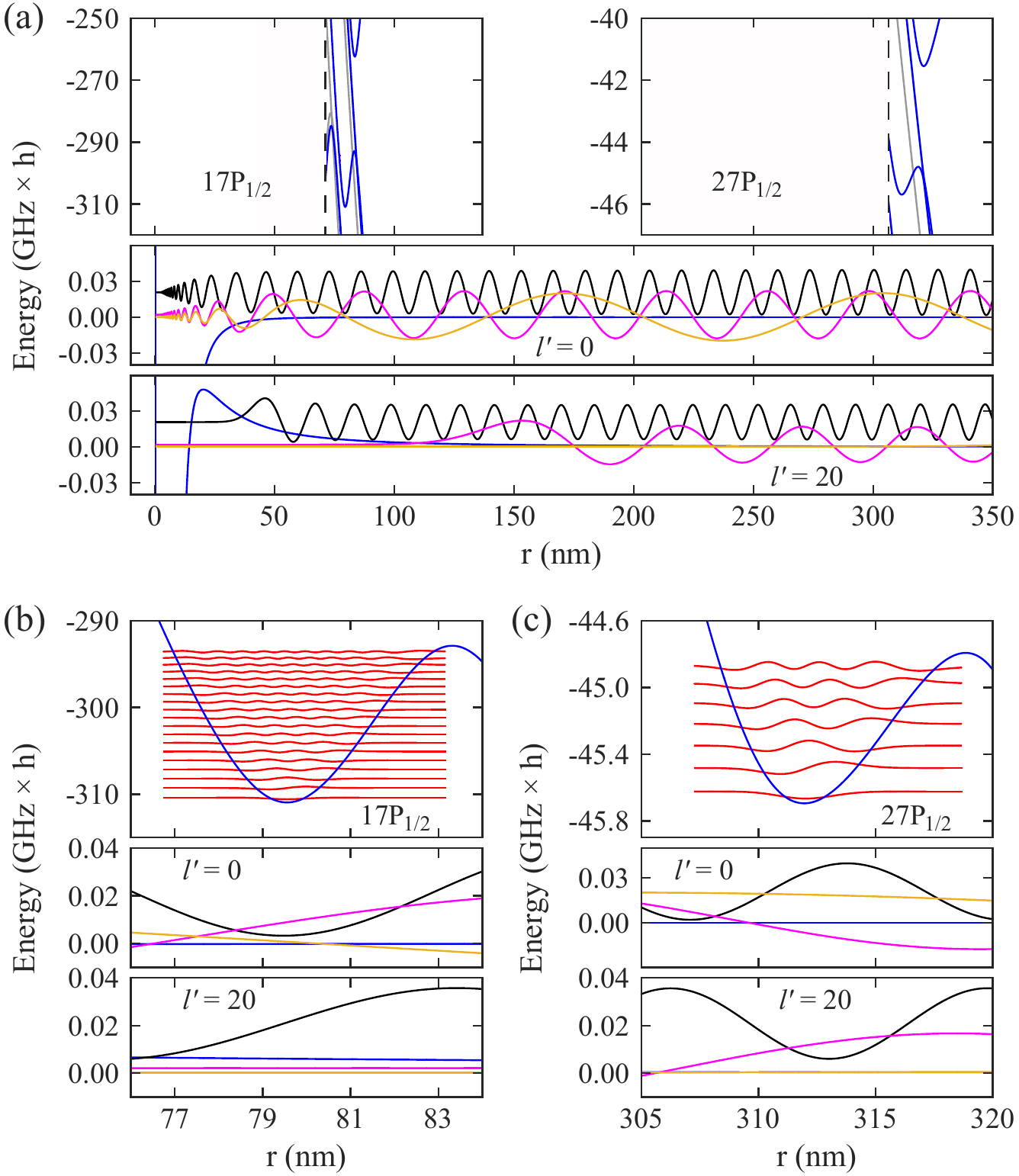}
	\caption{Comparison of wave functions. (\textbf{a}) Upper panel: Potential energy curves in the region of the two outermost potential wells associated with the $17P_{1/2}$ (on the left; see also Figure\ref{Fig1}b) and the $27P_{1/2}$ (on the right) states. Blue (gray) solid lines indicate levels with $|m_J|=1/2$ ($|m_J|=3/2$). Black dashed vertical lines mark internuclear distances below which the potential energy landscape is not plotted. The energy reference is the term energy of the $17P_{3/2}$ and $27P_{3/2}$ state, respectively, at zero electric field ($r\rightarrow \infty$). Middle panel: The blue solid lines show the potential energy curve for a Rb $5S_{1/2}$ atom colliding with a Ba$^+$ $6S_{1/2}$ ion within the partial wave $l'= 0$. We choose the $(1)^3\Sigma^+$ state to represent the short-range part for $r\lesssim 4\:\textrm{nm}$ (see text). Here, the energy reference corresponds to the atomic asymptote of the two collision partners. The black, magenta, and ocher solid lines are scattering wave functions for atom-ion collision energies $E$ of $(1,\,0.1,\,0.01)\:\textrm{mK}\times k_\mathrm{B}$, respectively. Their amplitudes are scaled for better visibility, and are given in arbitrary units. Lower panel: The same as the middle panel but for $l'=20$. For the blue solid lines the angular momentum potential is included. (\textbf{b} and \textbf{c}) Magnifications of parts of Figure~\ref{Fig4}a. For simplicity, in the upper panels only the second outermost potential wells associated with the states $17P_{1/2}$ and $27P_{1/2}$ are considered. The red solid lines are the wave functions of vibrational levels with $l'=0$ (presented by using a scaling factor and given in arbitrary units).
	\label{Fig4}
	}
\end{figure}

A possible way to create a long-range atom-ion Rydberg molecule is resonant photoassociation. For example, we consider a single ion immersed in a cloud of ultracold neutral Rb atoms in the ground state $5S_{1/2}$. When a colliding atom-ion pair reaches the distance of the molecular bond length ($\approx r_\mathrm{min}$ of a potential well) a laser with a wavelength of $\approx 300\:\textrm{nm}$ can resonantly drive a transition to an atomic Rydberg state with sufficient $P$ character. The partial wave will generally not change or only slightly change during photoassociation. Specifically, if atom and ion collide in a partial wave with angular momentum $l'$, then the produced atom-ion molecule will typically have a rotational angular momentum of $l'$ or $l'\pm1$ (see discussion further below). The binding lengths are quite large, ranging between $\approx 80\:\textrm{nm}$ and $\approx 1440\:\textrm{nm}$ in our examples. In particular, they are much larger than the typical internuclear distance where the maximum of the angular momentum barrier is located (see, e.g., the lower panel of Figure~\ref{Fig4}a). Therefore, even for relatively low collision energies of $1\:\textrm{mK}\times k_\mathrm{B}$, where $k_\mathrm{B}$ is Boltzmann's constant, quite a number of partial waves can contribute to the photoassociation. The precise number of partial waves depends of course on the atomic mass of the involved particles and the photoassociation distance, in addition to the precise collision energy (cf. Equation~\ref{Eq:lmax} further below). 

In Figure~\ref{Fig4} various examples for different parameters for photoassociation are presented. Specifically, here, we consider a $5S_{1/2}$ electronic ground state $^{87}$Rb atom colliding with a $^{138}$Ba$^+$ ion in its electronic ground state $6S_{1/2}$. For simplicity, the hyperfine structure of the Rb atom is ignored. Furthermore, in general, the total electronic spin degree of freedom is not taken into account, i.e. we do not discriminate between electronic singlet and triplet states. The blue solid lines in the middle and the lower panel of Figure~\ref{Fig4}a show the interaction potential for two different partial waves $l'=0$ (the $s$-wave) and $l'=20$. Regarding $l'=0$, we choose for very short range ($r\lesssim 4\:\textrm{nm}$) as interaction potential the $(1)^3\Sigma^+$ potential energy curve taken from \cite{Mohammadi2021}. The part at longer range ($r\gtrsim 4\:\textrm{nm}$) for $l'=0$ is given by the polarization potential $\propto 1/r^4$ (see, e.g., \cite{Haerter2014}). We have checked that these two parts are smoothly connected with each other. For $l' = 20$ the angular momentum potential of Equation~\ref{Eq:centrifugal}, which gives rise to the centrifugal barrier, is added. The black, magenta, and ocher solid lines in the middle and the lower panel of Figure~\ref{Fig4}a are calculated scattering wave functions for collision energies $E$ of $(1,\,0.1,\,0.01)\:\textrm{mK}\times k_\mathrm{B}$, respectively. These are energy-normalized (cf. Equation~\ref{Eq:scatwf} below). In order to carry out photoassociation it is important that the scattering wave function and the wave function of the target molecular level have sufficient Franck-Condon overlap. We present in the upper panel of Figure~\ref{Fig4}a the potential energy curves in the regions of the two outermost molecular potential wells associated with the $17P_{1/2}$ (on the left) and $27P_{1/2}$ (on the right) states of Rb. Figures~\ref{Fig4}b and \ref{Fig4}c are zooms into these regions. Here, for simplicity, in the upper panels we only show the second outermost potential wells together with the wave functions of the corresponding vibrational levels. A comparison to the scattering wave functions (middle and lower panels of Figures~\ref{Fig4}b and \ref{Fig4}c) reveals that these can have decent overlap with atom-ion Rydberg molecule states, in general. We note, however, that the lower the collision energy the smaller the number of partial waves which will contribute to the photoassociation. For example, this can be seen from the almost vanishing amplitude of the scattering wave function for the case $E=0.01\:\textrm{mK}\times k_\mathrm{B}$ and $l'=20$ over essentially the whole range of internuclear distances shown in Figure~\ref{Fig4}a.
	
Next, we provide a quantitative estimation for photoassociation. For low enough laser intensity the occupation probability $P_m$ of an atom-ion molecular state is given by

\begin{equation}
P_m =    n_\mathrm{at} \frac{\pi v_\mathrm{rel} }{k^2} \sum_{ l' = 0}^{l'_\mathrm{max}} (2 l' + 1)
 \frac{  \gamma_s}{\Delta^2 + (\gamma + \gamma_s)^2 }\,.
\label{eq:Pm}
\end{equation} 
This result is extracted from \cite{Jones2006} where an expression for the rate $\Gamma = P_m \gamma$ for excitation of the molecular state followed by its spontaneous decay is presented. Equation~\ref{eq:Pm} is valid only for small values of $P_m$. Here, $n_\mathrm{at}$ is the particle density of the atomic cloud, $v_\mathrm{rel}$ is the relative velocity between the collision partners, $k = \sqrt{2 \mu E / \hbar^2 } = \mu v_\mathrm{rel}/ \hbar $ is the wave number for reduced mass $\mu$, and $\Delta$ is the detuning from resonance. The highest contributing partial wave is roughly determined by

\begin{equation}
l'_\mathrm{max} \approx  \sqrt{ 2 \mu r_\mathrm{min}^2 \, E }/ \hbar \,.
\label{Eq:lmax}
\end{equation}
In Equation~\ref{eq:Pm}, $\gamma$ is the rate of natural spontaneous emission for the excited state, and $\gamma_s$ is the rate for stimulated decay back to the entrance channel. $\gamma_s$ can be expressed by

\begin{equation}
\hbar \gamma_s = 2 \pi  \left( \frac{\hbar \Omega_\mathrm{R}}{2} \right)^2   |\langle  \Psi_e | E,  l' \rangle|^2\,,
\end{equation} 
which involves the Franck-Condon overlap of the excited molecular bound state wave function $\Psi_e $ and the scattering wave function $| E,  l' \rangle$. Here, $\Omega_\mathrm{R}$ is the Rabi frequency for the optical coupling,

\begin{equation}
\hbar \Omega_\mathrm{R}    
   = - \mathcal{E}_0 d(r)  =  - \sqrt{ \frac{2 I}{c \, \varepsilon_0  } }   d(r) \,,
\label{Eq:coup}
\end{equation} 
where $I$ is the intensity of the light, $c$ is the speed of light, and $d(r)$ is the dipole matrix element for the optical transition. In Equation~\ref{Eq:coup}, $\mathcal{E}_0$ denotes the amplitude of the oscillating electric field of the light, which has a field strength of $\mathcal{E}_0 \cos(\omega t)$. We note that the scattering wave function $| E,  l' \rangle$ is energy normalized so that for $r \rightarrow \infty $ it takes the form

\begin{equation}
\langle r | E,  l' \rangle \approx \sqrt{ \frac{2\mu}{\pi \hbar^2 k} } \, \sin(kr - l' \frac{\pi}{2} + \delta_{ l'})\,.
\label{Eq:scatwf}
\end{equation} 
This energy normalization assures that the given wave has an incoming particle flux which is independent of $k$.
	
The transition electric dipole moment $d(r)$ varies with $n$, $r$, and the polarization of light (see, e.g., \cite{Wang2020}). For simplicity, we consider here the transition electric dipole moment $d_\infty$ for $r\rightarrow \infty$, which neglects Stark level shifts and mixing of states due to the electric field of the ion, in principle. Regarding transitions with $\pi$-polarized light from ($5S_{1/2}$, $m_J=1/2$) toward ($nP_{1/2}$, $m_J=1/2$), generally ignoring the hyperfine structure, one obtains that $|d_\infty|$ decreases from $5.5\times 10^{-3}ea_0$ for $n=17$ to $0.9\times 10^{-3}ea_0$ for $n=47$, where $a_0 = 0.529 \times 10^{-10}\:\textrm{m}$ is the Bohr radius. According to \cite{Wang2020}, $d(r)$ is on the order of $d_\infty/8$ for such transitions, which should be a reasonable approximation for our purpose. To give an example, we now determine the occupation probability $P_m$ as given in Equation~\ref{eq:Pm} for vibrational levels within the second outermost potential well associated with the $27P_{1/2}$ state of Rb. For this state a natural lifetime of $\tau=1/\gamma=3.9\times 10^{-5}\:\textrm{s}$ and $|d_\infty|/8=0.3\times 10^{-3}ea_0$ are calculated. We use an atomic density of $n_\mathrm{at}= 1\times 10^{12}\:\textrm{cm}^{-3}$, which is low enough such that atom-atom-ion three-body recombination is negligible during the typical duration of the photoassociation experiment (see, e.g., \cite{Kruekow2016}). Furthermore, an atom-ion collision energy of $E=1\:\textrm{mK}\times k_\mathrm{B}$ is considered. We assume that we resonantly address a molecular state with rotational angular momentum quantum number $l'_\mathrm{res}$ corresponding to $\Delta=0$. All other possible transitions between scattering states and molecular rotational states for a given vibrational level are accounted for with their respective detunings $\Delta \neq 0$. 
For a light intensity $I = 1\:\textrm{kW}\,\textrm{cm}^{-2}$  and $l'_\mathrm{res}=100$
we obtain from Equation~\ref{eq:Pm} an occupation probability $P_m$ on 
the order of one percent for some of the vibrational levels in the well, which 
is already sizable. 
By increasing the laser intensity the occupation probability $P_m$ still grows, but
in general not linearly anymore with the laser power. 
 We note that when choosing $l'_\mathrm{res}=10$ ($l'_\mathrm{res}=0$), $P_m$ is reduced typically   by about a factor of 8 (20) for the lot of vibrational levels as compared to $l'_\mathrm{res}=100$.
	
The vibrational levels on the potential wells should be spectroscopically resolvable when the collision energies of the atom-ion system are sufficiently low. For example, the energy level shift for a typical (thermally distributed) atom-ion collision energy of $1\:\textrm{mK}\times k_\mathrm{B}$ is about $21\:\textrm{MHz}\times h$, which entails a corresponding inhomogeneous line broadening. Thus, according to Figure~\ref{Fig2}e vibrational levels for the second outermost molecular potential wells associated with $nP_{1/2}$ Rb Rydberg states should be well resolvable for $n \lesssim 30$, as the vibrational level splitting is $\gtrsim 100\:\textrm{MHz}\times h$. The molecular rotation will not lead to significant additional broadening if the rotational angular momentum $l'$ does not change significantly in the photoassociation process. The maximal change in $|\Delta l'|$ due to recoil from the ultraviolet photon with wavelength $\lambda \approx 300\:\textrm{nm}$ in the Rydberg excitation can be estimated using $|\Delta l'| = (h/ \lambda) r_\mathrm{min}/ h = r_\mathrm{min}/ (300\:\textrm{nm})$. For $n = 30$ the approximate binding length $r_\mathrm{min}$ for molecules of the second outermost potential well is $\approx 400\:\textrm{nm}$, and hence one obtains $|\Delta l'| \le 1$. Furthermore, these parameters correspond to a rotational constant $B$ of about $0.6\:\textrm{kHz}\times h$ when considering a $^{87}$Rb$^{138}$Ba$^+$ molecule. According to Equation~\ref{Eq:lmax}, $l'_\mathrm{max}$ is $\approx 190$ for this system, assuming a collision energy of $E=1\:\textrm{mK}\times k_\mathrm{B}$. Consequently, about 190 partial waves can contribute to the photoassociation. The maximal energy shift (and line broadening) due to the change $\Delta l'$ is then on the order of $2B l'_\mathrm{max}|\Delta l'| \approx 2\:\textrm{MHz}\times h$, which is small as compared to the vibrational splitting of about $100\:\textrm{MHz}\times h$, and as compared to the thermal broadening for collision energies on the order of $1\:\textrm{mK}\times k_\mathrm{B}$.

\subsection{Detection by photoionization}	

In order to detect a long-range atom-ion Rydberg molecule, one can use photoionization. The resulting two positively charged ionic cores of the molecule repel each other such that the dimer gets dissociated. As a consequence, the total number of ionic particles increases, which can be observed with an ion detector. For heteronuclear long-range atom-ion Rydberg molecules mass-resolved ion detection could even discriminate whether a detected ion originated from the involved Rydberg atom. 
	
The efficiency of the detection scheme is directly related to the photoionization rate $\Gamma_\mathrm{PI}=\sigma_\mathrm{PI}I\lambda_\mathrm{PI}/(hc)$. Here, $\sigma_\mathrm{PI}$ represents the photoionization cross section and $\lambda_\mathrm{PI}$ is the wavelength of the photoionizing light. As an example, we consider the wavelength of $\lambda_\mathrm{PI}=1064\:\textrm{nm}$ for which high power laser sources are available. According to \cite{Cardman2021}, the photoionization cross section $\sigma_\mathrm{PI}$ for direct photoionization starting from Rb Rydberg $P$ states ranges from about $1\times10^{-21}\:\textrm{cm}^2$ for $n=90$ to about $2\times10^{-19}\:\textrm{cm}^2$ for $n=20$. Consequently, when addressing, e.g., the $17P$ state of Rb, a laser intensity on the order of $10^5\:\textrm{W}\,\textrm{cm}^{-2}$ is needed to obtain a photoionization rate comparable to the rate for natural radiative decay. Generally, the cross section approximately scales as $\propto n^{-3}$. Since, however, the lifetime of the molecule increases approximately as $\propto n_\mathrm{eff}^3$, there is only a comparatively small change of the ionization efficiency as a function of $n$ for a given laser intensity.
 
Additional information can be drawn from the  photoionization detection if it is combined with an energy spectroscopy of the released ions. Upon dissociation of the molecule each of the ionic fragments gains a characteristic kinetic energy determined by the electrostatic potential energy $E_\mathrm{pot} \approx e^2/(4\pi \varepsilon_0r_\mathrm{min})$, where $r_\mathrm{min}$ is roughly the binding length of the initial molecule. For $r_\mathrm{min}\approx 80\:\textrm{nm}$, as given for the potential wells involving $17P$ states considered in Figure~\ref{Fig1}b, one obtains a total kinetic energy of about $200\:\textrm{K}\times k_\mathrm{B}$ (corresponding to $\approx 17\:\textrm{meV}$), which is distributed over the two ionic fragments according to their mass ratio. Therefore, energy spectroscopy of product ions could reveal information about the binding length of initial long-range atom-ion Rydberg molecules. Such energy spectroscopy may be carried out, e.g., by employing an ion trap \cite{Mahdian2020} or by making use of time of flight techniques \cite{Mamyrin2001,Ashfold2006,Liu2020}.

\section{Prospects for experiments with long-range atom-ion Rydberg molecules}
	
\begin{figure}[t]
	\includegraphics[width=7.9cm]{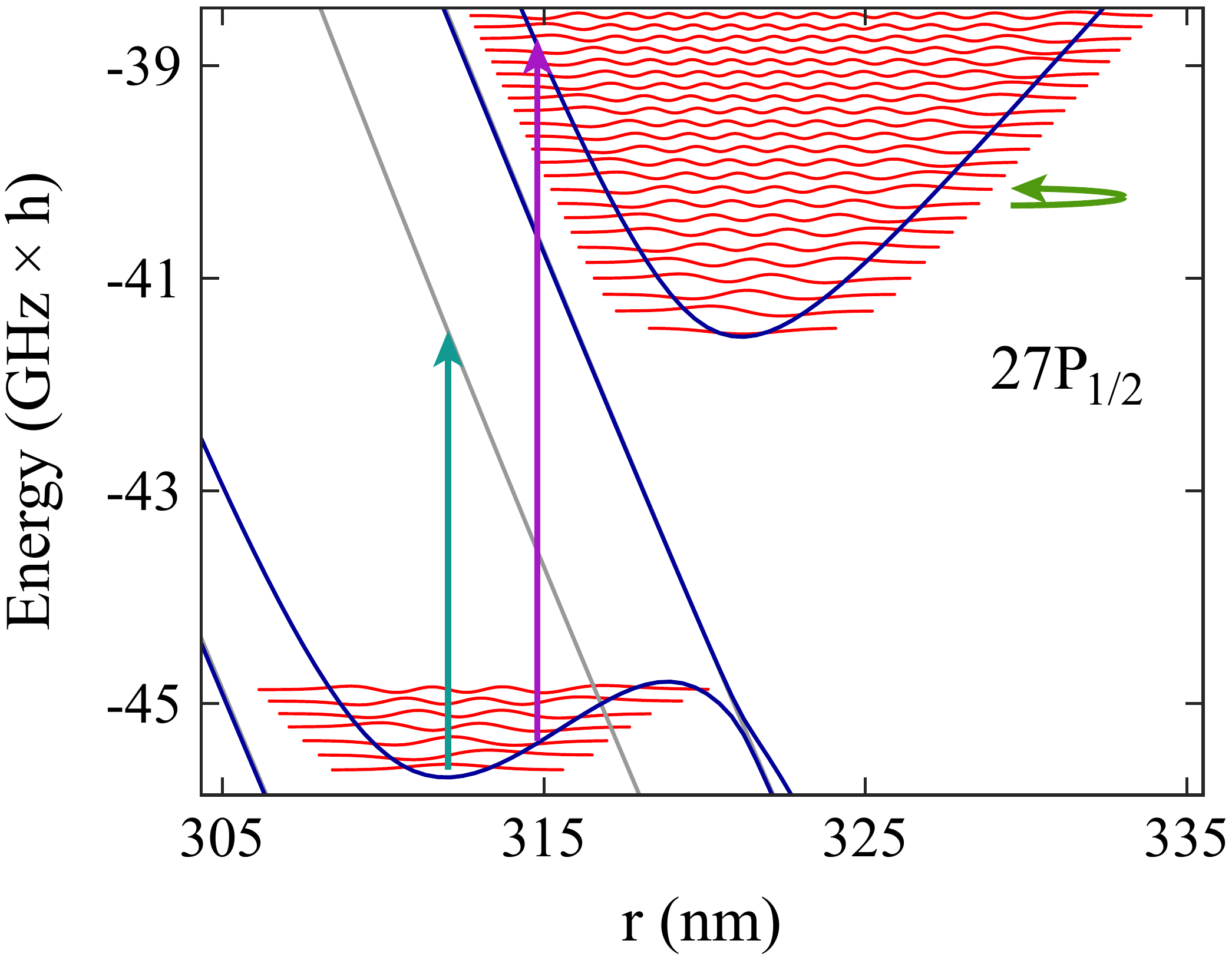}
	\caption{Examples for coupling of molecular levels. Shown are the potential energy curves in the region around the two outermost molecular potential wells associated with the $27P_{1/2}$ state of Rb. Blue (gray) solid lines indicate levels with $|m_J|=1/2$ ($|m_J|=3/2$). Here, the reference for zero energy is the term energy of the $27P_{3/2}$ state at zero electric field ($r \rightarrow \infty$). The red solid lines correspond to wave functions of vibrational levels with $l'=0$. Their amplitudes are scaled for better visibility, and are given in arbitrary units. The green arrow illustrates a transition between two neighboring vibrational states within the same potential well, which can be driven by a radio frequency field. The purple vertical arrow represents a microwave transition between two vibrational states in different potential wells. The cyan vertical arrow indicates a transition from a bound state towards a repulsive potential energy curve. \label{Fig5}
	}
\end{figure}

The long-range atom-ion Rydberg molecules possess several unusual properties which make them appealing for future investigations and applications. For example, they exhibit a very large permanent electric dipole moment with respect to their barycenter, even for the homonuclear case. Considering, e.g., a homonuclear dimer with an approximate binding length of $r_\mathrm{min}\approx 80\:\textrm{nm}$ (see also Figure~\ref{Fig1}b), the calculated dipole moment is approximately $er_\mathrm{min}/2 \approx6.4\times 10^{-27}\:\textrm{Cm} \approx 1900\:\textrm{D}$. Since the rotational constants are very small, already small external electric fields on the $\textrm{V}/\textrm{m}$ scale or below will be sufficient to align long-range atom-ion Rydberg molecules and to perform related experiments with them. Furthermore, their vibrational oscillation frequencies are in the $10\:\textrm{MHz}$ to $1\:\textrm{GHz}$ regime for the range of states considered here (cf. Figure~\ref{Fig2}e). The corresponding timescale of $1$ to $100\:\textrm{ns}$ is convenient for the investigation of wave packet dynamics, since it is largely accessible using standard lab electronics and laser switches. Microwave and radio frequency radiation can be employed to drive transitions between vibrational and rotational states (see Figure~\ref{Fig5} for different examples). This gives rise to interesting opportunities for coupling various neighboring potential wells in order to form novel, complex potential landscapes. The wave packet dynamics in such potential landscapes, which might involve tunneling and non-adiabatic transitions between states, can then be studied in detail. Related methods for potential engineering using, e.g., magnetic or electric fields are currently developed for (neutral) long-range Rydberg molecules \cite{Hollerith2021, Hummel2021}.

\section{Conclusions and outlook}
\label{Outlook}

We have predicted a novel molecular binding mechanism between an atomic ion and a neutral Rydberg atom. The electric field of the ion induces Stark shifts and level crossings in the Rydberg atom which gives rise to potential wells for long-range atom-ion Rydberg molecules. These molecules can possess extremely large binding lengths on the micrometer scale and correspondingly large dipole moments in the kilo-Debye range. From our investigation we expect the stability and lifetime to be sufficient for detection and further applications. We have characterized the properties of the molecule as a function of the principal quantum number of the involved Rydberg atom, also regarding the vibrational and rotational level structure. In addition, we have proposed methods for production and detection of the long-range atom-ion Rydberg molecule, and discussed prospects for studies of wave packet dynamics and potential landscape engineering. 

One possible platform for the observation of the long-range atom-ion Rydberg molecule are hybrid atom-ion experiments. In these experiments either laser-cooled and trapped ions are immersed into ultracold trapped clouds of neutral atoms, or ions are directly produced within an ultracold parent gas (for reviews, see, e.g., \cite{Haerter2014,Tomza2017}). Typical setups allow for a high level of control over the collision energy between atom and ion, for which values around $1\:\textrm{mK}\times k_\mathrm{B}$ and below have been achieved \cite{Mohammadi2019, Dieterle2021, Feldker2020, Schmidt2020}. We note that in Paul traps, electric fields are used to confine ions. These are, however, comparatively small. Typically, the electric field strength even at a distance of a few $\upmu\textrm{m}$ away from the trap center is well below $100\:\textrm{V}/\textrm{m}$. Binding lengths in the range from $1.0$ to $0.1\:\upmu\textrm{m}$ for a long-range atom-ion Rydberg molecule already correspond to ion electric fields at the position of the neutral atom in the range from $1.44$ to $144\:\textrm{kV}/\textrm{m}$. Therefore, the additional external electric field due to the ion trap only leads to a small distortion of the relevant molecular potential wells. Considering the current status of the field of hybrid atom-ion experiments, we anticipate that an observation of the long-range atom-ion Rydberg molecule proposed here is well within reach. 
 
During preparation of the manuscript we became aware of parallel work to ours \cite{Duspayev2021}.

\vspace{6pt} 


\authorcontributions{
J.H.D and M.D developed the concepts. S.H. carried out calculations of Stark maps, scattering wave functions and the occupation probability. M.D. and J.H.D. wrote the manuscript. J.H.D. supervised the project.
}
	
\funding{This research was funded by the German Research Foundation (DFG) within the priority program "Giant Interactions in Rydberg Systems" (DFG SPP 1929 GiRyd).
}

\acknowledgments{We thank Dominik Dorer for comments, and Guido Pupillo, N\'{o}ra S\'{a}ndor, and Georg Raithel for discussions on  collisions between a Rydberg atom and an ion.}

\conflictsofinterest{The authors declare no conflict of interest.}
	
\end{paracol}
\reftitle{References}

%


\end{document}